\documentclass{iau}
\usepackage{graphicx}

\title[A confined flare above filaments] 
{A confined flare above filaments}

\author[K. Dalmasse \& R. Chandra \& B. Schmieder \& G. Aulanier]   
{K. Dalmasse$^1$,
 R. Chandra$^2$,
 B. Schmieder$^1$,
 \and G. Aulanier$^1$}

\affiliation{$^1$LESIA, Observatoire de Paris, CNRS, UMPC, Univ. Paris Diderot, \\
5 place Jules Janssen, 92190 Meudon, France \\ email: {\tt kevin.dalmasse@obspm.fr} \\[\affilskip]
$^2$Dept. of Physics, DSB Campus, Kumaun University, Nainital- 263 002, India 
}

\pubyear{2008}
\volume{xxx}  
\pagerange{119--126}
\jname{Title of your IAU Symposium}
\editors{A.C. Editor, B.D. Editor \& C.E. Editor, eds.}
\begin{document}

\maketitle

\begin{abstract}
We present the dynamics of two filaments and a C-class flare observed in 
NOAA 11589 on 2012 October 16. We used the multi-wavelength high-resolution 
data from SDO, as well as THEMIS and ARIES 
ground-based observations. The observations show that the filaments are 
progressively converging towards each other without merging. We find that the 
filaments have opposite chirality which may have prevented 
them from merging. On October 16, a C3.3 class flare occurred 
without the eruption of the filaments. According to the 
standard solar flare model, after the reconnection, post-flare loops form {\it below} 
the erupting filaments whether the eruption fails or not. However, the observations 
show the formation of post-flare loops {\it above} the filaments, which is not consistent 
with the standard flare model. We analyze the topology of the active region's magnetic 
field by computing the quasi-separatrix layers (QSLs) using a linear force-free field extrapolation. 
We find a good agreement between the photospheric footprints of the QSLs and the 
flare ribbons. We discuss how slipping or slip-running reconnection 
at the QSLs may explain the observed dynamics.
\end{abstract}

\firstsection 
              
\section{Introduction}

Filaments are dark, elongated structures consisting of chromospheric plasma 
embedded in the much hotter corona 
(\cite[van Ballegooijen \& Martens 1989]{vanBallegooijen89}; 
\cite[Chae \etal\ 2001]{Chae01}). They are cool 
($\approx 8000 \ \textrm{K}$), dense material confined in highly stressed 
magnetic fields overlying polarity inversion lines (PILs; 
\cite[Aulanier \& D\'emoulin 1998]{Aulanier98}; 
\cite[Schmieder \etal\ 2006]{Schmieder06}). 
In the standard picture, the magnetic structure of filaments is formed through 
shearing motions along PILs and/or, through magnetic flux cancellation due 
to converging motions of opposite magnetic polarities towards the PILs 
(\cite[van Ballegooijen \& Martens 1989]{vanBallegooijen89}; 
 \cite[Antiochos \etal\ 1994]{Antiochos94}).

Eventually, filaments may become unstable 
(see \cite[Moore \etal\ 2001]{Moore01}; \cite[Martens \& Zwaan 2001]{Martens01}). 
According to the 
standard solar flare model (hereafter, CSHKP model), the instability of 
the filament may lead to two different types of flares, namely, eruptive or 
confined flares (see review by \cite[Shibata \& Magara 2011]{Shibata11}). 
Eruptive flares correspond to cases for which the filament erupts, 
leading to the formation of a coronal mass ejection 
(CME). On the other hand, confined flares are sometimes associated with 
cases for which the eruption of the filament fails 
({\it e.g.}, \cite[T\"or\"ok \& Kliem 2005]{Torok05}). 
Confined flares also comprise flares induced by magnetic 
reconnection of different magnetic flux tubes, or magnetic coronal loops, 
for which no filament is present 
({\it e.g.}, \cite[Berlicki \etal\ 2004]{Berlicki04}; \cite[Chandra \etal\ 2006]{Chandra06}). 
In the context of flares involving the presence of a filament, the CSHKP model 
predicts for both eruptive and confined flares, that the flare will be associated 
with two flare ribbons, and with the formation of hot post-flare loops {\it below} 
the erupting filament, regardless of whether it is a successful or failed eruption.

In this study, we present the evolution of two filaments and a confined flare 
observed in NOAA 11589, which cannot be explained by 
the CSHKP model. We propose an alternative flare scenario which accounts 
for the observed flare signatures and filaments evolution during the flare.

\section{Observations}

Our study was performed by combining observations from the Solar Dynamic 
Observatory (SDO) satellite, the french T\'elescope H\'eliographique 
pour l'Etude du Magn\'etisme et des Instabilit\'es Solaires (THEMIS), 
and an indian telescope of the Aryabhatta Research Institute of observational 
Sciences (ARIES).

NOAA 11589 appeared on 2012 October 10 at the heliographic coordinates 
N13 E61. 
The AR quickly developed into two decaying magnetic polarities 
(see Fig.\,\ref{Fig-Bz-AIA171}a). 
During its on-disk passage, the AR was associated with large-scale magnetic 
flux cancellation, and a few localized magnetic flux emergence events.

The flux cancellation in the internal part of the AR 
led to the formation of two filaments of opposite chirality which eventually 
converged. However, the filaments did not merge probably due to their axial field 
being oriented in opposite direction along the PIL 
({\it e.g.}, \cite[Schmieder \etal\ 2004]{Schmieder04}; \cite[DeVore \etal\ 2005]{DeVore05}).

The AR also presented some recurring and localized magnetic flux emergence 
associated with Ellerman bombs (EBs) in its northern part 
(as in \cite[Pariat \etal\ 2004]{Pariat04}).

On October 16, the AR produced a confined C3.3 class flare which started around 
16:00 UT and ended around 16:39 UT. A first analysis of the flare signatures with 
AIA 1600 \AA\, and AIA 171 \AA\, channels seem to be in agreement with the CSHKP model: 
apparently presenting two flare ribbons, and the formation of hot post-flare loops.

However, a careful analysis of the EUV data from the AIA 171 \AA\, channel reveals that 
the flare did not lead to the eruption of any of the filaments. It also shows 
a striking result: the post-flare loops were formed {\it above} the filaments 
contrary to what is expected from the CSHKP model, and the filaments were not 
disturbed by the flare.

\section{Analysis}

\subsection{Magnetic field extrapolation}

To understand and explain the evolution of the filaments during the flare, we study 
the magnetic topology of the AR by means of an LFFF extrapolation 
($\vec{\nabla} \times  \vec{B}= \alpha \vec{B}$, with $\alpha$ being the force-free parameter) 
to identify the key sites for the development 
of magnetic reconnection that led to the flare. 

We only considered 
the global magnetic field of the AR because (i) the filaments were in 
plage regions where the magnetic field is weak, and thus, the currents are 
not well measured, and (ii) the filaments did not seem to play 
any role in the flare.

\begin{figure}[h]
\begin{center}
 \includegraphics[width=0.99\textwidth]{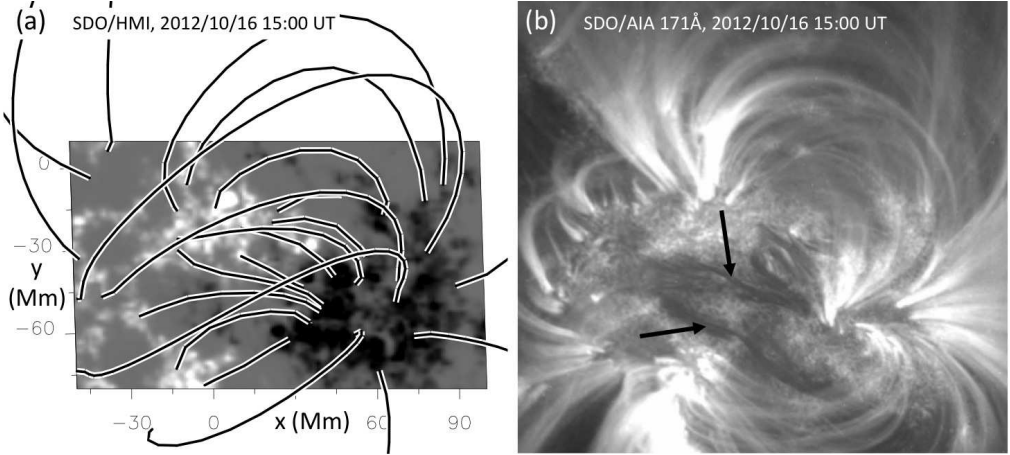} 
 \caption{Central part of NOAA 11589. (a) Photospheric vertical magnetic field, $B_z$, in greyscale overplotted with selected magnetic field lines from the extrapolation (black lines). (b) AIA171 image showing some of the AR loops and the two observed filaments highlighted by black arrows.}
   \label{Fig-Bz-AIA171}
\end{center}
\end{figure}

The extrapolations were performed using the fast Fourier transform method 
(\cite[Alissandrakis 1981]{Alissandrakis81}) with a non-uniform 
grid of $1024^2 \times 351$ points covering a domain of 
$700^2 \times 2000 \ \textrm{Mm}^3$. Within the set of performed extrapolations, 
we kept the solution $\alpha = 7 \times 10^{-3} \ \textrm{Mm}^{-1}$ because it gave 
the best match with the northern loops of the AR (Fig.\,\ref{Fig-Bz-AIA171}a), 
{\it i.e.}, the region where the flare was initiated.

\subsection{Topological analysis}

The topology is then analyzed by computing the quasi-separatrix layers 
(QSLs; {\it e.g.}, \cite[D\'emoulin \etal\ 1996]{Demoulin96}). 
QSLs are thin 3D volumes of very sharp gradients of the magnetic field connectivity. 
QSLs are preferential sites for the build-up of strong and thin current layers, 
and for the development of magnetic reconnection at these current layers 
(see review by \cite[D\'emoulin 2006]{Demoulin2006}). As separatrices, QSLs are preferential 
sites for particle acceleration (\cite[Aulanier \etal\ 2006]{Aulanier06}). 
Many observational studies have thus successfully 
compared and associated the photospheric footprints of QSLs to flare ribbons providing 
indirect evidence of magnetic reconnection as the triggering mechanism of solar 
eruptive events ({\it e.g.}, \cite[D\'emoulin \etal\ 1996]{Demoulin96}; 
\cite[Mandrini \etal\ 1997]{Mandrini97}; \cite[Schmieder \etal\ 1997]{Schmieder97}). 
The photospheric mapping of QSLs can be obtained by 
computing the squashing degree, $Q$ (\cite[Titov \etal\ 2002]{Titov02}). 
QSLs are thus identified as 3D regions of strong 
$Q$-values ($Q \gg 2$).

\begin{figure}[h]
\begin{center}
 \includegraphics[width=0.99\textwidth]{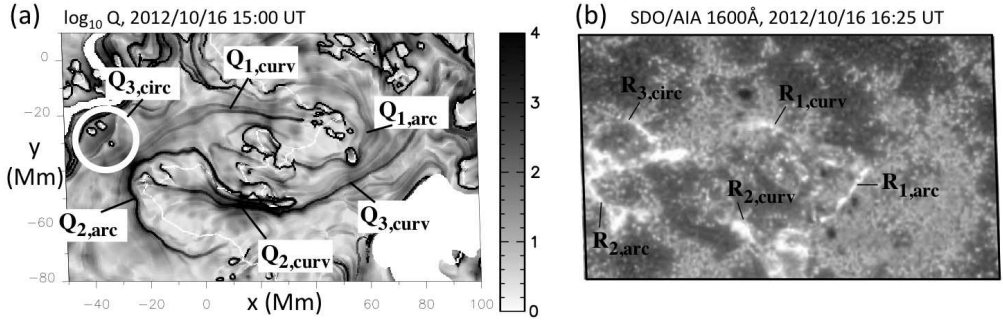} 
 \caption{Central part of NOAA 11589. (a) Photospheric mapping of $\log_{10} Q$ displaying the photospheric footprints of QSLs at 15:00 UT. (b) Flare ribbons at 16:25 UT. The footprints of three QSLs, labelled $Q_{i}$, are identified with the three flare ribbons labelled $R_{i}$.}
   \label{Fig-QSLs-AIA1600}
\end{center}
\end{figure}

We computed the squashing degree for our LFFF extrapolation using ``method 3'' 
of \cite{Pariat12}. 
Fig.\,\ref{Fig-QSLs-AIA1600}(a) displays the photospheric mapping of QSLs for the same 
field of view as Fig.\,\ref{Fig-Bz-AIA171}(a) by representing $\log_{10} Q$ at the photosphere. 
By plotting magnetic field lines over the photospheric $Q$-map, we identified two double-C 
shape QSLs, $Q_{1,2}$, similar to \cite{Aulanier05}, and a circular-like one 
(overlaid with a white circle), $Q_3$, similar to \cite{Masson09}. 
We find a few discrepancies between the QSLs footprints, $Q_{i}$, and the three 
flare ribbons of Fig.\,\ref{Fig-QSLs-AIA1600}(b), $R_{i}$. This is due 
to the assumptions made by extrapolating the AR's magnetic field in LFFF, 
which do not model the highly-stressed filament magnetic fields, and which 
results in local modifications of the magnetic connectivity that slightly 
modifies the location and shape of the QSLs in our extrapolation. 
Nevertheless, there is a good qualitative agreement between the QSLs footprints and 
the flare ribbons (Fig.\,\ref{Fig-QSLs-AIA1600}).

\section{Conclusion}

From the previous analysis, it is clear that the magnetic field of AR 11589 presents 
a complex topology formed by three entangled QSLs. 
Such a complex topology was favorable 
to the build-up of electric current layers and to the development of magnetic 
reconnection at any of these QSLs. The flare might thus have been 
the result of the stress of, at least, one of the QSLs eventually triggering 
magnetic reconnection at all QSLs.

Analyzing the AIA and HMI data prior to, and after the flare, we found signatures 
of localized, recurring magnetic flux emergence in the northern part of the AR --- 
in the region below $Q_1$, {\it i.e.}, between the western part of $Q_{1,curv}$ 
and the southern part of $Q_{1,arc}$. 
Consequently, we propose that this episodic flux emergence was 
the driver of the C3.3 class flare: 
this continuous magnetic flux emergence may have 
stressed the magnetic field of $Q_1$, resulting in the formation of a strong thin 
current layer, at least, within this QSL. Eventually, this can 
trigger slipping or slip-running reconnection at $Q_1$ 
(see \cite[Aulanier \etal\ 2006]{Aulanier06}), which, 
in turn, can trigger magnetic reconnection at all the other intersecting QSLs, $Q_2$ 
and $Q_3$. 
This would have induced particle acceleration at all QSLs 
({\it e.g.}, \cite[Masson \etal\ 2009]{Masson09}), and hence, the formation of 
a complex distribution of flare ribbons (such as shown by 
Fig.\,\ref{Fig-QSLs-AIA1600}b). Since both filaments were located below 
the QSLs involved in the flare mechanism, our scenario naturally accounts 
for the development of post-flare loops {\it above} these non-erupting filaments.


\end{document}